\documentclass[aps,showpacs,twocolumn]{revtex4}
\usepackage{epsfig}

\begin{document}

\title{The structure of pentaquarks $P_c^+$ in the chiral quark model}

\author{Gang Yang$^1$, Jialun Ping$^1$}

\affiliation{Department of Physics and Jiangsu Key Laboratory for Numerical
Simulation of Large Scale Complex Systems, Nanjing Normal University, Nanjing 210023, P. R. China}

\begin{abstract}
  The recent experimental results of LHCb collaboration suggested the existence of pentaquark states with charmonium.
  To understand the structure of the states, a dynamical calculation of 5-quark systems with quantum numbers
  $IJ^P=\frac{1}{2}(\frac{1}{2})^\pm$, $\frac{1}{2}(\frac{3}{2})^\pm$ and $\frac{1}{2}(\frac{5}{2})^\pm$ is performed
  in the framework of chiral quark model with the help of gaussian expansion method. The results show that the negative
  parity states can be bound states while all of the positive parity states are the scattering states.
  The $P_c(4380)$ state is suggested to be the bound state of $\Sigma^*_cD$. Although the energy of $\Sigma_cD^*$ is
  very close to the mass of $P_c(4450)$, the inconsistent parity prevents the assignment. The calculated distances 
  between quarks confirm the molecular nature of the states. Other five-quark bound states of the combination of 
  $\Sigma_cD$ and $\Sigma^*_cD^*$  are also found in the region about 4.3GeV and 4.5GeV. 
\end{abstract}


\maketitle

\section{Introduction}

Since the report of $\Theta^+(1540)$ by several groups~\cite{M1,M2,M3} about ten years ago, it brought lots of arguments
during that time. Although the observation of pentaquark state $\Theta^+(1540)$ was not confirmed by the further
experiments~\cite{notheta} (LEPS Collaboration still insisted on the existence of pentaquark $\Theta^+(1540)$~\cite{theta2}),
the study of pentaquark inspires new structures for hadrons, beyond the conventional quark configuration (${qqq}$
or ${q}\bar{q}$). Five-quark components in baryons was also studied which showed that the $qqqq\bar{q}$ in ground
state is more favorable than $qqq$ with $L=1$ for $1/2^-$ baryons~\cite{M4}.

Recently, the interesting in pentaquark is revived, because LHCb experiment reported the observation of two pentaquark
states, denoted as $P_c^+(4380)$ and $P_c^+(4450)$, in the decay of $\Lambda_b^0$, $\Lambda_b^0 \rightarrow
J/\psi K^-p$~\cite{M6}. The masses and widths of these two structures, appeared in the $J/\psi p$ invariant mass,
are determined to be 4380$\pm8\pm29$ MeV, 205$\pm18\pm86$ MeV, and 4449.8$\pm1.7\pm2.5$ MeV, 39$\pm5\pm19$ MeV. The pentaquark nature
of the structures comes from the valence structure, $uudc\bar{c}$, of $J/\psi p$. The possible quantum numbers $J^P$ of
these two states are $(3/2^-, 5/2^+)$ or $(5/2^-, 3/2^+)$. In fact, the hidden-charm pentaquark states have been
predicted several years ago. In 2010, J. J. Wu {\em et al.} predicted several narrow resonances with hidden charm above 4 GeV,
$N^*_{c\bar{c}}(4265)$, $N^*_{c\bar{c}}(4415)$, and $\Lambda^*_{c\bar{c}}(4210)$, in the framework of the coupled-channel
unitary approach~\cite{PRL105}. Z. C. Yang {\em et al.} also studied the possible existence of very loosely bound
hidden-charm molecular baryons in
the one-boson-exchange model, $\Sigma_c\bar{D}^*$ and $\Sigma_c\bar{D}$ states are proposed~\cite{CPC36}. After the report
of LHCb, a lot of theoretical are devoted to explain the nature of the two states. By using the boson exchange model, R. Chen
{\em et al.} interpreted the two states as the molecular states, $\Sigma_c(2455)D^*$ and $\Sigma^*_c(2520)D^*$ with
spin-parity $J^P=3/2^-$ and $J^P=5/2^-$, respectively~\cite{RChen}. The Bethe-Salpeter equation method was employed to
study the $\bar{D}\Sigma^*$ and $\bar{D}^*\Sigma_c$ interactions, and the two states $P_c^+(4380)$ and $P_c^+(4450)$
are identified as $\Sigma^*_c\bar{D}$ and $\Sigma_c(2455)\bar{D}^*$ molecular states with quantum numbers $J^P=3/2^-$
and $J^P=5/2^+$, respectively~\cite{JHe}. In QCD sum rule approach, $P_c^+(4380)$ and $P_c^+(4450)$ were explained as
hidden-charm pentaquark states with quantum numbers $J^P=3/2^-$ and $J^P=5/2^+$, respectively, by using
diquark-diquark-antiquark type interpolating currents~\cite{HXChen,ZGWang}. By analyzing the reaction
$\Lambda^0_{b}\to J/\psi$$K^-p$ with coupled-channel calculation, L. Roca {\em et al.} assigned the quantum numbers
$J^P=3/2^-$ to the state $P_c^+(4450)$ and concluded that $P_c(4450)^+$ state is a molecular state of mostly
$\Sigma_c\bar{D}^*$ and $\Sigma^*_c\bar{D}^*$ with $3/2^-$~\cite{Roca}. In the soliton approach, the hidden-charm
state with quantum numbers $IJ^P=\frac{1}{2}{\frac{3}{2}}^-$ was shown to exist and is compatible with $P_c^+(4380)$,
but the state with $IJ^P=\frac{1}{2}{\frac{5}{2}}^+$ has much higher mass compared with that of $P_c^+(4450)$~\cite{soliton}.
The small mass splitting between $P_c^+(4380)$ and $P_c^+(4450)$ can be understood in the diquark-triquark model
by using an effective diquark-triquark Hamiltonian based on spin-orbital interaction~\cite{ZQ}. Non-resonance
explanations of the structures observed experimentally were also proposed~\cite{ATS,PRD92}.

Based on the theory of QCD, it is possible to excite quark-antiquark pairs from vacuum to form hadronic state.
For the light quark-antiquark pair excitation, the effect can be absorbed into the parameters in the quark model
description. For the heavy quark-antiquark pair excitation, it is too difficult to occur in the light hadron system
and its effect cannot be absorbed by the model parameters. So the states $P_c^+$ reported by LHCb should be genuine
pentaquarks. Its study will provide us more information of the underlying fundamental theory of strong interaction,
QCD.

The most common approach to multiquark system is quark model. After fifty years development and with the accumulation of
experimental data on multiquark states, to tackle the problem of multiquark seriously in the framework of quark model is
expected. In the present work, the chiral quark model is used to study the pentaquark states with hidden-charm.
Different from other approaches, no prior spacial structure of the state is assumed, the structure is determined by
the system dynamics. For this purpose, a powerful method of few-body system, gaussian expansion mthod (GEM)~\cite{GEM}
is employed to do the calculation. The GEM has been successfully applied to many few-body systems, light nuclei,
hypernuclei, hadron physics and so on~\cite{GEM}. It suits for both of compact multi-quark systems and loosely bound
molecular states.

The structure of the paper is as follows. In section II the quark model, wavefunctions and calculation method is presented.
Section III is devoted to the calculated results and discussions. A brief summary is given in the last section.

\section{model and wave function}

The chiral quark model has acquired great achievement both in describing the hadron spectra and hadron-hadron interaction.
Here we apply it to 5-quark system. The details of the model can be found in Ref.\cite{ChQM}. The Hamiltonian for multiquark
system takes the form
\begin{widetext}
\begin{eqnarray}
H & = & \sum_{i=1}^{n}\left( m_i+\frac{p^2_i}{2m_i}\right)-T_{CM}
      + \sum_{j>i=1}^{n}\left[ V_{CON}({{\bf r}_{ij}})+V_{OGE}({{\bf r}_{ij}})
      +V_{\chi}({{\bf r}_{ij}})+V_{\sigma}({{\bf r}_{ij}})\right] , \\
V_{CON}({{\bf r}_{ij}}) & = & \mbox{\boldmath $\lambda$}_i^c\cdot\mbox{\boldmath $\lambda$}_j^c
   \left[-a_c (1-e^{-\mu_cr_{ij}})+\Delta \right] , \nonumber \\
V_{OGE}({{\bf r}_{ij}}) & = & \frac{1}{4}\alpha_s\mbox{\boldmath $\lambda$}_i^c\cdot\mbox{\boldmath $\lambda$}_j^c
   \left[ \frac{1}{r_{ij}}-\frac{1}{6m_im_j}\mbox{\boldmath $\sigma$}_i\cdot\mbox{\boldmath $\sigma$}_j
   \frac{e^{-r_{ij}/r_0(\mu)}}{r_{ij}r^2_0(\mu)}\right] , ~~~r_0(\mu)=\hat{r}_0/\mu,~~\alpha_{s} =
   \frac{\alpha_{0}}{\ln(\frac{\mu^2+\mu_{0}^2}{\Lambda_{0}^2})}. \nonumber \\
V_{\sigma}({{\bf r}_{ij}}) & = & -\frac{g^2_{ch}}{4\pi} \frac{\Lambda^2_\sigma}{\Lambda^2_\sigma-m^2_\sigma}m_\sigma
	\left[ Y(m_{\sigma}r_{ij})-\frac{\Lambda_\sigma}{m_\sigma}Y(\Lambda_{\sigma}r_{ij})\right] ,\\
V_{\chi}({{\bf r}_{ij}}) & = & v_{\pi}({{\bf r}_{ij}})\sum_{a=1}^{3}
	(\mbox{\boldmath $\lambda$}_i^a\cdot\mbox{\boldmath $\lambda$}_j^a)+v_{K}({{\bf r}_{ij}})\sum_{a=4}^{7}
	(\mbox{\boldmath $\lambda$}_i^a\cdot\mbox{\boldmath $\lambda$}_j^a)+v_{\eta}({{\bf r}_{ij}})
    [\cos\theta_{P}(\mbox{\boldmath $\lambda$}_i^8\cdot\mbox{\boldmath $\lambda$}_j^8)-sin\theta_{P}] , \nonumber \\
v_{\chi}({{\bf r}_{ij}}) & = & \frac{g^2_{ch}}{4\pi}\frac{m^2_\chi}{12m_im_j}
	\frac{\Lambda^2_\chi}{\Lambda^2_\chi-m^2_\chi}m_\chi
	\left[ Y(m_{\chi}r_{ij})-\frac{\Lambda^3_\chi}{m^3_\chi}Y (\Lambda_{\chi}r_{ij}) \right]
	(\mbox{\boldmath $\sigma$}_i\cdot\mbox{\boldmath $\sigma$}_j).    ~~~~\chi=\pi,K,\eta
\end{eqnarray}
\end{widetext}
Here $T_{CM}$ is the center of mass kinetic energy, $\mu$ is the reduced mass of two interacting quark pair.
Because we are interested in the ground state of multiquark system,
only the central part of the interaction is given above. All the symbols take their usual meanings.
The model parameters of the model are taken from Ref.\cite{ChQM} and are listed in Table I.
\begin{table}[h]
\caption{Quark model parameters. The masses of mesons take their experimental mass, i.e.,
$m_{\pi}=0.70$ fm$^{-1}$, $m_\sigma=3.42$ fm$^{-1}$, $m_\eta=2.77$ fm$^{-1}$,
$m_K=2.51$ fm$^{-1}$. \label{model}}
\begin{tabular}{c|cc cc} \hline
  Quark masses     &$m_u$=$m_d\;(MeV)$   &313 \\
                   &$m_c\;(MeV)$  &1752\\   \hline
                   &$\Lambda_\pi=\Lambda_\sigma~$ (fm$^{-1}$)  &4.20\\
  Goldstone bosons &$\Lambda_\eta=\Lambda_K~$ (fm$^{-1}$)      &5.20\\
                   &$g^2_{ch}/(4\pi)$  &0.54\\
                   &$\theta_P(^\circ)$  &-15\\  \hline
            &$a_c$ (MeV)  &430\\
Confinement &$\mu_c$ (fm$^{-1})$  &0.70\\
            &$\Delta$ (MeV)  &181.10\\
            &$\alpha_s$  &0.777\\ \hline
       &$\alpha_0$  &2.118\\
       &$\Lambda_0~$(fm$^{-1}$)  &0.113\\
OGE    &$\mu_0~$(MeV)  &36.976\\
       &$\hat{r}_0~$(MeV~fm)  &28.170\\
       &$\hat{r}_g~$(MeV~fm)  &34.500\\ \hline
\end{tabular}
\end{table}

The wavefunctions for the system are constructed in the following way.
First the 5-quark system is separated into two clusters, one with 3 quarks
and another with quark-antiquark. The wavefunctions for these sub-clusters
can be easily written down. Then two clusters are coupled and anti-symmetrized
(if necessary) to form the total wavefunction of 5-quark system. Clearly, there
are other ways to construct the wavefunctions of the system. However, it makes
no difference by choosing any one configuration that if we use enough bases
for system during the calculation.

For the 5-quark system with quark content $uudc\bar{c}$, there are types of separation,
one is $(udc)(\bar{c}u)+(uuc)(\bar{c}d)$ and the other is $(uud)(c\bar{c})$. Due to the
large mass difference between $c$ quark and $u$, $d$ and $s$ quarks, the flavor SU(4)
symmetry is strongly broken. Here we just construct the flavor wavefunctions of system
based on the flavor SU(2) symmetry. The flavor wavefunctions for the sub-clusters
constructed are shown below.
\begin{eqnarray}
&& |B_{11}\rangle =uuc,~~  |B_{10}\rangle = \frac{1}{\sqrt{2}}(ud+du)c,
~~|B_{1-1}\rangle = ddc, \nonumber \\
&& |B_{00}\rangle = \frac{1}{\sqrt{2}}(ud-du)c, \nonumber \\
&&|B_{\frac12,\frac12}^1\rangle = \frac{1}{\sqrt{6}}(2uud-udu-duu),  \\
&&|B_{\frac12,\frac12}^2\rangle = \frac{1}{\sqrt{2}}(ud-du)u, \nonumber \\
&&|M_{\frac12,\frac12}\rangle = \bar{c}u, ~~~~
|M_{\frac12,-\frac12}\rangle = \bar{c}d, ~~~~
|M_{00}\rangle = \bar{c}c. \nonumber
\end{eqnarray}
The flavor wavefunctions for 5-quark system with isospin $I=1/2$ are obtained by
the following couplings,
\begin{eqnarray}
|\chi^f_1 \rangle & = & \sqrt{\frac{2}{3}}|B_{11}\rangle |M_{\frac12,-\frac12}\rangle
 -\sqrt{\frac{1}{3}}|B_{10}\rangle |M_{\frac12,\frac12}\rangle, \nonumber \\
|\chi^f_2 \rangle & = & |B_{00}\rangle |M_{\frac12,\frac12}\rangle, \\
|\chi^f_3 \rangle & = & |B_{\frac12,\frac12}^1 \rangle |M_{00}\rangle, \nonumber \\
|\chi^f_4 \rangle & = & |B_{\frac12,\frac12}^2 \rangle |M_{00}\rangle, \nonumber
\end{eqnarray}

In a similar way, the spin wavefunctions for 5-quark system can be constructed,
\begin{eqnarray}
& & |\chi_{\frac12,\frac12}^{\sigma 1}(5)\rangle = \sqrt{\frac{1}{6}}
|\chi_{\frac32,-\frac12}^{\sigma}(3)\rangle |\chi_{11}^{\sigma}\rangle
-\sqrt{\frac{1}{3}} |\chi_{\frac32,\frac12}^{\sigma}(3)\rangle |\chi_{10}^{\sigma}\rangle
\nonumber \\
&& ~~~~~~~~~~~~+\sqrt{\frac{1}{2}} |\chi_{\frac32,\frac32}^{\sigma}(3)\rangle |\chi_{1-1}^{\sigma}\rangle \nonumber \\
& & |\chi_{\frac12,\frac12}^{\sigma 2}(5)\rangle = \sqrt{\frac{1}{3}}
|\chi_{\frac12,\frac12}^{\sigma 1}(3)\rangle |\chi_{10}^{\sigma}\rangle
-\sqrt{\frac{2}{3}} |\chi_{\frac12,-\frac12}^{\sigma 1}(3)\rangle |\chi_{11}^{\sigma}\rangle  \nonumber \\
& & |\chi_{\frac12,\frac12}^{\sigma 3}(5)\rangle = \sqrt{\frac{1}{3}}
|\chi_{\frac12,\frac12}^{\sigma 2}(3)\rangle |\chi_{10}^{\sigma}\rangle
-\sqrt{\frac{2}{3}} |\chi_{\frac12,-\frac12}^{\sigma 2}(3)\rangle |\chi_{11}^{\sigma}\rangle  \nonumber \\
& & |\chi_{\frac12,\frac12}^{\sigma 4}(5)\rangle = |\chi_{\frac12,\frac12}^{\sigma 1}(3)\rangle
  |\chi_{00}^{\sigma}\rangle   \nonumber \\
& & |\chi_{\frac12,\frac12}^{\sigma 5}(5)\rangle = |\chi_{\frac12,\frac12}^{\sigma 2}(3)\rangle
  |\chi_{00}^{\sigma}\rangle   \\
& & |\chi_{\frac32,\frac32}^{\sigma 1}(5)\rangle = \sqrt{\frac{3}{5}}
|\chi_{\frac32,\frac32}^{\sigma}(3)\rangle
  |\chi_{10}^{\sigma}\rangle -\sqrt{\frac{2}{5}} |\chi_{\frac32,\frac12}^{\sigma}(3)\rangle
  |\chi_{11}^{\sigma}\rangle \nonumber \\
& & |\chi_{\frac32,\frac32}^{\sigma 2}(5)\rangle = |\chi_{\frac32,\frac32}^{\sigma}(3)\rangle
  |\chi_{00}^{\sigma}\rangle  \nonumber \\
& & |\chi_{\frac32,\frac32}^{\sigma 3}(5)\rangle = |\chi_{\frac12,\frac12}^{\sigma 1}(3)\rangle
  |\chi_{11}^{\sigma}\rangle  \nonumber \\
& & |\chi_{\frac32,\frac32}^{\sigma 4}(5)\rangle = |\chi_{\frac12,\frac12}^{\sigma 2}(3)\rangle
  |\chi_{11}^{\sigma}\rangle  \nonumber \\
& & |\chi_{\frac52,\frac52}^{\sigma 1}(5)\rangle = |\chi_{\frac32,\frac32}^{\sigma}(3)\rangle
  |\chi_{11}^{\sigma}\rangle  \nonumber
\end{eqnarray}
with the spin wavefunctions for 3-quark and 2-quark sub-clusters,
\begin{eqnarray}
&& |\chi_{\frac32,\frac32}^{\sigma}(3)\rangle =\alpha\alpha\alpha,~~ \nonumber \\
&&   |\chi_{\frac32,\frac12}^{\sigma}(3)\rangle = \frac{1}{\sqrt{3}}
   (\alpha\alpha\beta+\alpha\beta\alpha+\beta\alpha\alpha), \nonumber \\
&&   |\chi_{\frac32,-\frac12}^{\sigma}(3)\rangle = \frac{1}{\sqrt{3}}
   (\alpha\beta\beta+\beta\alpha\beta+\beta\beta\alpha),  \nonumber \\
&& |\chi_{\frac12,\frac12}^{\sigma 1}(3)\rangle =\frac{1}{\sqrt{6}}
   (2\alpha\alpha\beta-\alpha\beta\alpha-\beta\alpha\alpha), \nonumber \\
&& |\chi_{\frac12,\frac12}^{\sigma 2}(3)\rangle =\frac{1}{\sqrt{2}}
   (\alpha\beta\alpha-\beta\alpha\alpha), \nonumber \\
&& |\chi_{\frac12,-\frac12}^{\sigma 1}(3)\rangle =\frac{1}{\sqrt{6}}
   (\alpha\beta\beta-\alpha\beta\beta-2\beta\beta\alpha), \nonumber \\
&& |\chi_{\frac12,-\frac12}^{\sigma 2}(3)\rangle =\frac{1}{\sqrt{2}}
   (\alpha\beta\beta-\beta\alpha\beta), \nonumber \\
&& |\chi_{11}^{\sigma}\rangle =\alpha\alpha,~~
   |\chi_{10}^{\sigma}\rangle =\frac{1}{\sqrt{2}}
   (\alpha\beta+\beta\alpha),~~|\chi_{1-1}^{\sigma}\rangle =\beta\beta, \nonumber \\
&& |\chi_{00}^{\sigma}\rangle =\frac{1}{\sqrt{2}} (\alpha\beta-\beta\alpha). \nonumber
\end{eqnarray}

For the color wavefunction, only the color singlet channel, two clusters are all colorless,
is used here. The reason for this simplification comes from that the color singlet channel
states are complete when all the excitation of other degrees of freedom are included
for the multiquark systems, the energies for the excited states are rather high and have
small effect on the group states. Then the color wavefunction of the system is
\begin{equation}
\chi^c=\frac{1}{\sqrt{6}}(rgb-rbg+gbr-grb+brg-bgr)\frac{1}{\sqrt{3}}(\bar r r+\bar gg+\bar bb).
\end{equation}

As for the orbital wavefunctions, we do not separate the motions of particles in the system
into internal and relative ones and freeze the internal motion, the structure of the system
is assumed priorly, as the most work did. In the present work, the orbital wavefunctions for
each motion of the system are determined by the dynamics of the system, so does the structure.
Another reason for not including the hidden-color channels in the calculation is that the
direct extension of interactions between quark pairs from colorless states to colorful states
are questionable, it will lead to too much bound states~\cite{JPG39}.
The orbital wavefunctions for this purpose is obtained as follows,
\begin{equation}
\psi_{LM_L}=\left[ \left[ \left[
  \phi_{n_1l_1}(\mbox{\boldmath $\rho$})\phi_{n_2l_2}(\mbox{\boldmath $\lambda$})\right]_{l}
  \phi_{n_3l_3}(\mbox{\boldmath $r$}) \right]_{l^{\prime}}
  \phi_{n_4l_4}(\mbox{\boldmath $R$}) \right]_{LM_L}
\end{equation}
where the Jacobi coordinates are defined as,
\begin{eqnarray}
{\mbox{\boldmath $\rho$}} & = & {\mbox{\boldmath $x$}}_1-{\mbox{\boldmath $x$}}_2, \nonumber \\
{\mbox{\boldmath $\lambda$}} & = & {\mbox{\boldmath $x$}}_3
 -(\frac{{m_1\mbox{\boldmath $x$}}_1+{m_2\mbox{\boldmath $x$}}_2}{m_1+m_2}),  \\
{\mbox{\boldmath $r$}} & = & {\mbox{\boldmath $x$}}_4-{\mbox{\boldmath $x$}}_5, \nonumber \\
{\mbox{\boldmath $R$}} & = & (\frac{{m_1\mbox{\boldmath $x$}}_1+{m_2\mbox{\boldmath $x$}}_2
  +{m_3\mbox{\boldmath $x$}}_3}{m_1+m_2+m_3})
  -(\frac{{m_4\mbox{\boldmath $x$}}_4+{m_5\mbox{\boldmath $x$}}_5}{m_4+m_5}). \nonumber
\end{eqnarray}
To find the orbital wavefunctions, the Gaussian expansion method (GEM) is employed, i.e.,
each $\phi$ is expanded by gaussians with various sizes~\cite{GEM}
\begin{equation}
 \phi_{nlm}(\mbox{\boldmath $r$})=\sum_{n=1}^{n_{max}} c_n N_{nl}r^le^{-(r/r_n)^2}Y_{lm}(\hat{\mbox{\boldmath $r$}}),
\end{equation}
where $N_{nl}$ is normalization constants
\begin{equation}
 N_{nl}=\left[\frac{2^{l+2}(2\nu_n)^{l+\frac{3}{2}}}{\sqrt \pi(2l+1)}\right]^{\frac{1}{2}}.
\end{equation}
The size parameters $r_n$ are taken as the geometric progression numbers
\begin{equation}
 \nu_n=1/r^2_n~~~~ r_n=r_{min}a^{n-1}.
\end{equation}
$c_n$ is the variational parameters, which is determined by the dynamics of the system.
Finally, the complete channel wave function for the 5-quark system is written as
\begin{equation}
 \Psi_{JM,i,j,n}={\cal A} \left[ \left[
   \chi^{\sigma_i}_{S}(5) \psi_{L}\right]_{JM_J}
   \chi^{f}_j \chi^{c} \right]
\end{equation}
where the $\cal{A}$ is the antisymmetry operator of the system, it can be written as
\begin{equation}
 {\cal{A}}=1-(15)-(25)
\end{equation}
for $(udc)(\bar cu)$ case and
\begin{equation}
 {\cal{A}}=1-(13)-(23)
\end{equation}
for $(uud)(c\bar c)$ case.

The eigen-energy of the system is obtained by solving the following eigen-equation
\begin{equation}
H\Psi_{JM}=E\Psi_{JM},
\end{equation}
by using variational principle. The eigen functions $\Psi_{JM}$ are the linear combination of the
above channel wavefunctions.

When the angular momenta are not all zero, the calculation of the matrix elements of Hamiltonian
is rather complicated. Here a new useful method named the infinitesimally-shifted Gaussian (ISG)
are used~\cite{GEM}. In this method, the orbital wavefunctions are written as
\begin{equation}
  \phi_{nlm}(\mbox{\boldmath $r$})=N_{nl}\lim_{\varepsilon\to 0}\frac{1}{(\nu \varepsilon)^l}
  \sum_{k=1}^{k_{max}}C_{lm,k}e^{{-\nu_n (\mbox{\boldmath $r$}-\varepsilon \mbox{\boldmath $D$}_{lm,k})}^2}
\end{equation}
the coefficients $C_{lm,k}$ and shift-direction vector $\mbox{\boldmath $D$}_{lm,k}$ are
dimensionless numbers. By absorbing the spherical harmonic function into the shifted gaussians,
the calculation becomes easy with no tedious angular-momentum algebra required.

\section{Results and discussions}

In the present calculation, we are interested in the low-lying states of $uudc\bar{c}$
pentaquark system, so the total orbital angular momentum $L$ is limited to be 0 and 1.
For $L=0$, all of $l_1,l_2,l_3,l_4$ are 0 and for $L=1$, only one of $l_1,l_2,l_3,l_4$
can be 1. In this way, the total angular momentum $J$ can take values 1/2, 3/2 and 5/2.
The possible channels under the consideration are listed in Table II. The single channel
and channel coupling calculations are performed in this work. The results are shown in
Tables \ref{Gresult}. The tables gives the eigen-energies of the states (column 3),
along with the theoretical (column 4) and experimental thresholds (column 6), the binding
energies (column 5) and the corrected energies of the states (column 7), which are obtained
by taking the sum of experimental thresholds and the binding energies. Table \ref{Distance} gives the
spacial configurations of the states. In the following we analyse the results in detail.
\begin{table}[htb]
\caption{The channels under the consideration.  \label{channels}}
\begin{tabular}{cl} \hline
  $J^P$        & ~~~$|LM_L;SM_S\rangle$ \\ \hline
  $\frac{1}{2}^-$    & ~~~$|00;\frac{1}{2}\frac{1}{2}\rangle$ \\  \hline
  $\frac{1}{2}^+$    & ~~~$\sqrt{\frac{2}{3}}$$|11;\frac{1}{2}\,$-$\frac{1}{2}\rangle
                       -\sqrt{\frac{1}{3}}|10;\frac{1}{2}\frac{1}{2}\rangle$ \\
                     & ~~~$\sqrt{\frac{1}{6}}$$|11;\frac{3}{2}\,$-$\frac{1}{2}\rangle
                       -\sqrt{\frac{1}{3}}|10;\frac{3}{2}\frac{1}{2}\rangle
                       +\sqrt{\frac{1}{2}}|1\,$-$1;\frac{3}{2}\frac{3}{2}\rangle$ \\  \hline
  $\frac{3}{2}^-$    & ~~~$|00;\frac{3}{2}\frac{3}{2}\rangle$ \\  \hline
  $\frac{3}{2}^+$    & ~~~$|11;\frac{1}{2}\frac{1}{2}\rangle$ \\
                     & ~~~$\sqrt{\frac{2}{5}}$$|11;\frac{3}{2}\frac{1}{2}\rangle
                       -\sqrt{\frac{3}{5}}$$|10;\frac{3}{2}\frac{3}{2}\rangle$ \\
                     & ~~~$\sqrt{\frac{1}{15}}$$|11;\frac{5}{2}\frac{1}{2}\rangle
                       -\sqrt{\frac{4}{15}}$$|10;\frac{5}{2}\frac{3}{2}\rangle
                       +\sqrt{\frac{10}{15}}$$|1\,$-$1;\frac{5}{2}\frac{5}{2}\rangle$ \\  \hline
  $\frac{5}{2}^-$    & ~~~$|00;\;\frac{5}{2}\frac{5}{2}\rangle$ \\  \hline
  $\frac{5}{2}^+$    & ~~~$|11;\frac{3}{2}\frac{3}{2}\rangle$ \\ \hline
\end{tabular}
\end{table}

For the parity negative states, the results are shown in Table \ref{Gresult}.
\begin{table*}[htb]
\caption{Eigen-energies of the $udc{\bar{c}}u$ system with parity negative (unit: MeV).  \label{Gresult}}
\begin{tabular}{llccccc} \hline
  $J^{P}$  & Channel   & Eigen Energy  & E$_{th}$ (Theo.)  & Binding Energy &
    E$_{th}$ (Exp.)  & Corrected Energy \\ \hline
  ${\frac{1}{2}}^{-}$
    & $\chi^{\sigma i}_{1/2}\chi^f_j~~i=4,5,~~j=3,4$  & 3745 & 3745 & 0    & 3919($N\eta_c$)       & 3919 \\
    & $\chi^{\sigma i}_{1/2}\chi^f_j~~i=2,3,~~j=3,4$  & 3841 & 3841 & 0    & 4036($NJ/\psi$)       & 4036 \\
    & $\chi^{\sigma i}_{1/2}\chi^f_j~~i=4,5,~~j=2$  & 3996 & 3996 & 0    & 4151($\Lambda_cD$)    & 4151 \\
    & $\chi^{\sigma i}_{1/2}\chi^f_j~~i=2,3,~~j=2$  & 4115 & 4115 & 0    & 4293($\Lambda_cD^*$)  & 4293 \\
    & $\chi^{\sigma i}_{1/2}\chi^f_j~~i=4,5,~~j=1$  & 4398 & 4402 & $-4$ & 4320($\Sigma_cD$)     & 4316 \\
    & $\chi^{\sigma i}_{1/2}\chi^f_j~~i=2,3,~~j=1$  & 4518 & 4521 & $-3$ & 4462($\Sigma_cD^*$)   & 4459 \\
    & $\chi^{\sigma i}_{1/2}\chi^f_j~~i=1,~~j=1$  & 4563 & 4566 & $-3$ & 4527($\Sigma^*_cD^*$) & 4524 \\
    &  $~~~~~~mixed$ & 4397 & & & \\  \hline
  ${\frac{3}{2}}^{-}$
    & $\chi^{\sigma i}_{3/2}\chi^f_j~~i=3,4,~~j=3,4$  & 3841 & 3841 & 0    & 4036($NJ/\psi$)       & 4036 \\
    & $\chi^{\sigma i}_{3/2}\chi^f_j~~i=3,4,~~j=2$  & 4115 & 4115 & 0    & 4293($\Lambda_cD^*$)   & 4293 \\
    & $\chi^{\sigma i}_{3/2}\chi^f_j~~i=3,4,~~j=1$  & 4518 & 4520 & $-2$ & 4462($\Sigma_cD^*$)    & 4460 \\
    & $\chi^{\sigma i}_{3/2}\chi^f_j~~i=2,~~j=1$  & 4444 & 4447 & $-3$ & 4385($\Sigma^*_cD$)   & 4382 \\
    & $\chi^{\sigma i}_{3/2}\chi^f_j~~i=1,~~j=1$  & 4564 & 4566 & $-2$ & 4527($\Sigma^*_cD^*$)  & 4525 \\
    & $~~~~~~mixed$ & 4442 & & & & \\  \hline
  ${\frac{5}{2}}^{-}$
    & $\chi^{\sigma i}_{5/2}\chi^f_j~~i=1,~~j=1$  & 4563 & 4566 & $-3$ & 4527($\Sigma^*_cD^*$)  & 4524 \\ \hline
\end{tabular}
\end{table*}

(a) $J^P=\frac{1}{2}^-$: For $N\eta_c$, $NJ/\psi$, $\Lambda_c D$ and $\Lambda_c D^*$ states,
the single-channel calculation shows that no bound state can be formed. For $\Sigma_c D$,
$\Sigma_c D*$ and $\Sigma_c^* D$ states, bound states with $-3\sim -4$ MeV biding energies appear.
The channel-coupling is rather weak, it does not push the state $N\eta_c$ or $\Lambda_c D$ down
enough to form a bound state, and it also does not push the state $\Sigma_c D$ up above the threshold.
So the present calculation shows that there is a resonance $\Sigma_c D$ with resonance
energy 4315 MeV for $J^P=\frac{1}{2}^-$. The result is in agreement with that of
Ref.\cite{CPC36}. Although the energy of $\Sigma_c D^*$ is very close to that of
$P_c(4450)$, it is difficult to make the assignment because of the different parities.

(b) $J^P=\frac{3}{2}^-$: The similar results with the case of $J^P=\frac{1}{2}^-$
are obtained. $NJ/\psi$, $\Lambda_c D^*$ states are unbound and all $\Sigma_c D$'s are
bound states. The channel-coupling pushes down the state $\Sigma_c^* D$ a little.
So a resonance, $\Sigma_c^* D$, is shown up. After the correction, the resonance
energy is 4382 MeV, which is very close the mass of $P_c^+(4380)$, which was
claimed by LHCb collaboration~\cite{M1}. However, the large decay width of $P_c^+(4380)$
cannot be explained in the present calculation. The decay width of $\Sigma_c^* D$ state
to $NJ/\psi$, $\Lambda_cD^*$ are estimated to several MeVs due to the weak channel-coupling.
Because of the missing of the spin-orbit
interaction in the present calculation, the energies of $NJ/\psi$, $\Sigma_c D^*$ with
$J^P=\frac{3}{2}^-$ are the same as that of $\Sigma_c D^*$, $NJ/\psi$ with $J^P=\frac{1}{2}^-$.

(c) $J^P=\frac{5}{2}^-$: Only one channel: $\Sigma^*_c D^*$, remains in this case
if all orbital angular momenta are set to zero. Bound state is obtained as before.
Although the binding energy is small, $-3$ MeV, the decay width of the state may be
small, $10\sim 20$ MeV, due to the small decay widths of its constituents,
$\Sigma_c^*~(\Gamma_{\Sigma_c^*\to \Lambda_c\pi}\sim 15$ MeV) and
$D^* (\Gamma_{D^*\to D\pi}\sim 1$ MeV). So it is a good candidate of the
heavy pentaquark with high spin.

For the parity positive states, one of the orbital angular momenta is 1,
Generally it is difficult to form a bound state in this case. Our calculation
shows that all the states under investigation are not bounded.
So if the state $P_c^+(4450)$ is identified as a pentaquark state with positive
parity, the chiral quark model may be needed to be modified for multi-quark system.
The non-resonance explanation of the narrow structure at 4.45 GeV was also proposed,
Guo {\em et al.} showed that the structure was compatible with the kinematical effects
of the rescattering from $\chi_{c1}p$ to $J/\psi p$~\cite{PRD92}.

\begin{table*}[htb]
\caption{Distances between any two quarks (unit: fm).  \label{Distance}}
\begin{tabular}{llccccccc} \hline
  $J^{P}$  & ~~~~~~~~Channel   & ~~$r_{12}$~~ & ~~$r_{13}$~~ & ~~$r_{14}$~~ & ~~$r_{15}$~~
  & ~~$r_{34}$~~ & ~~$r_{35}$~~ & ~~$r_{45}$~~ \\ \hline
 ${\frac{1}{2}}^{-}$
  & $\chi^{\sigma i}_{1/2}\chi^f_j~~i=4,5,~~j=1$ ($\Sigma_cD$)   & 0.8  &0.7  &1.5 &1.6 &1.9 &1.8 &0.6\\
  & $\chi^{\sigma i}_{1/2}\chi^f_j~~i=2,3,~~j=1$ ($\Sigma_cD^*$) & 0.8   &0.7  &1.7 &1.6 &2.1 &1.8 &0.7\\
  & $\chi^{\sigma i}_{1/2}\chi^f_j~~i=1,~~j=1$ ($\Sigma_c^*D^*$) & 0.9  &0.8  &1.6 &1.6 &2.0 &1.7 &0.7\\
 \hline\hline
 ${\frac{3}{2}}^{-}$
  & $\chi^{\sigma i}_{3/2}\chi^f_j~~i=3,4,~~j=1$ ($\Sigma_cD^*$) & 0.8   & 0.7  &1.9 &1.7 &2.3 &2.0 &0.7\\
  & $\chi^{\sigma i}_{3/2}\chi^f_j~~i=2,~~j=1$ ($\Sigma_cD^*$) & 0.9  & 0.8  &1.5 &1.8 &2.1 &2.0 &0.6\\
  & $\chi^{\sigma i}_{3/2}\chi^f_j~~i=1,~~j=1$ ($\Sigma_c^*D^*$) & 0.9  & 0.8  &2.1 &2.1 &2.7 &2.3 &0.7\\
 \hline\hline
 ${\frac{5}{2}}^{-}$
  & $\chi^{\sigma i}_{5/2}\chi^f_j~~i=1,~~j=1$ ($\Sigma_c^*D^*$) & 0.9  & 0.8 &1.9 &1.8 &2.4 &2.0 &0.7\\ \hline
\end{tabular}
\end{table*}

To find the structure of the resonances obtained in the present work, the distances
between any two quarks are calculated. The results are shown in Table \ref{Distance}.
From the table, we can see that the distances among quarks 1, 2 and 3 is around 0.8 fm,
and the distances between quark 4 and antiquark are 0.6-0.7 fm, while the distances
between quarks 1, 2, 3 and 4, 5 are 1.5-2.4 fm. Clearly, quark 1,2,3 form a cluster,
$\Sigma(^*)$ and quark 4,5 form another cluster, $D(^*)$, then two clusters combine to
a pentaquark state. To describe $P_c^+(4380)$ as molecular state of $\Sigma^*_c D$ is reasonable.

Fig.~1 shows the correlation functions of two clusters for $J^P={\frac{3}{2}}^-$. Typical
behavior of the wavefunction for bound state is obtained.

\begin{figure}[ht]
\epsfxsize=3.5in \epsfbox{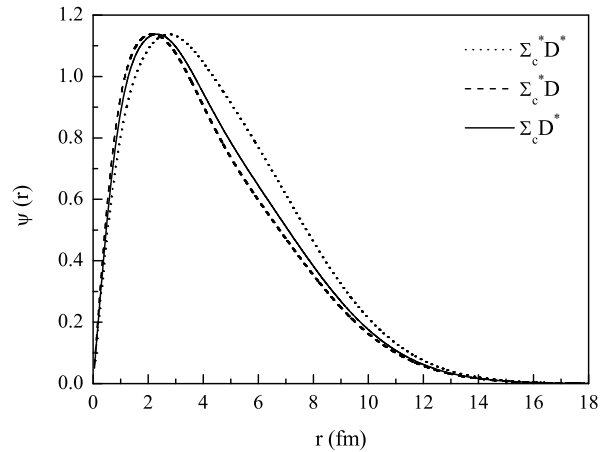} \vspace{-1.0cm}
\caption{The correlation functions of $\Sigma_cD^*$, $\Sigma^*_c D$ and $\Sigma^*_c D^*$
with $IJ^P=\frac{1}{2}{\frac{3}{2}}^-$.}
\end{figure}

\section{Summary}
In the framework of chiral quark model, the 5-quark systems with quark contents
$uddc\bar{c}$ are investigated by means of Gaussian expansion method.
The calculation shows that there are several resonances for
$IJ^P=\frac{1}{2}{\frac{3}{2}}^-,~\frac{1}{2}{\frac{3}{2}}^-$ and $\frac{1}{2}{\frac{5}{2}}^-$,
in which the mass of state $\frac{1}{2}{\frac{3}{2}}^-$ with configuration
$\Sigma_c^* D$ is very close to that of state $P_c^+(4380))$, a pentaquark
announced by LHCb collaboration. The distances between quark pairs suggest
a molecular structure for these resonances. A sound interpretation of $P_c^+(4380)$
is the molecule of $\Sigma_c^* D$ with $IJ^P=\frac{1}{2}{\frac{3}{2}}^-$. However,
the large decay width of $P_c^+(4380))$, 205$\pm18\pm86$ MeV, is out of
reach of the present picture. The mass of molecule state $\Sigma_c D^*$ with
$IJ^P=\frac{1}{2}{\frac{1}{2}}^-$ or $\frac{1}{2}{\frac{3}{2}}^-$ is also
close to that of $P_c^+(4450)$, another pentaquark reported by LHCb collaboration.
Nevertheless, the opposite parity of the state to the $P_c^+(4380)$ may prevent this
assignment. Meanwhile all the positive parity states are all unbound in our calculation.

In the present calculation, the spacial structure of a 5-quark system is not
assumed in advance. Although the two colorless sub-clusters are used, the internal
structures of the sub-clusters are not fixed. Generally the state in this approach
will have smaller energy than it in other approaches, because of the larger space.

As a preliminary work, the spin-orbit and tensor interactions are not included
in the calculation. For parity negative states, their effects are expected to
be zero or small. For parity positive state, it will play a minor role.
To understand the nature of $P_c^+(4450)$ in quark model approach, the calculation
with including the spin-orbit interaction is needed, which in progress in our group.

\section*{Acknowledgments}
The work is supported partly by the National Natural Science Foundation of China under Grant
Nos. 11175088, 11535005 and 11205091.

\end{document}